\documentclass{moriond}

\def\be{\begin{equation}}
\def\ee{\end{equation}}
\def\bea{\begin{eqnarray}}
\def\eea{\end{eqnarray}}

\usepackage[separate-uncertainty=true]{siunitx}
\DeclareSIUnit\gauss{G}

\newcommand{\Photo}{}

\begin{document}
\vspace*{4cm}
\title{TESTING THE GRAVITATIONAL REDSHIFT WITH GALILEO SATELLITES}

\author{ P. Delva$^{1}$, N. Puchades$^{2,1}$, E. Sch\"onemann$^3$, F. Dilssner$^3$, C. Courde$^4$, S. Bertone$^5$, F. Gonzalez$^6$, A. Hees$^1$, Ch. Le Poncin-Lafitte$^{1}$, F. Meynadier$^{1}$, R. Prieto-Cerdeira$^6$, B. Sohet$^{1}$, J. Ventura-Traveset$^7$, P. Wolf$^{1}$}

\address{$^1$SYRTE, Observatoire de Paris, Universit\'e PSL, CNRS, Sorbonne Universit\'e, LNE, 61 avenue de l'Observatoire 75014 Paris France}
\address{$^2$Departamento de Astronomia y Astrofisica - Valencia University}
\address{$^3$European Space Operations Center, ESA/ESOC, Darmstadt Germany}
\address{$^4$UMR Geoazur, Universit\'e de Nice, Observatoire de la C\^ote d'Azur, 250 rue A. Einstein, F-06560 Valbonne, France}
\address{$^5$Astronomical Institute, University of Bern, Sidlerstrasse 5 CH-3012 Bern, Switzerland}
\address{$^6$European Space and Technology Centre, ESA/ESTEC, Noordwijk, The Netherlands}
\address{$^7$European Space and Astronomy Center, ESA/ESAC, Villanueva de la Ca\~nada, Spain}

\maketitle\abstracts{
We present the results of the analysis of the GREAT (Galileo gravitational Redshift test with Eccentric sATellites) experiment from SYRTE (Observatoire de Paris), funded by the European Space Agency. An elliptic orbit induces a periodic modulation of the fractional frequency difference between a ground clock and the satellite clock, while the good stability of Galileo clocks allows to test this periodic modulation to a high level of accuracy. Galileo satellites GSAT0201 and GSAT0202, with their large eccentricity and on-board H-maser clocks, are perfect candidates to perform this test. By analyzing 1008 days of eccentric Galileo satellites data we measure the fractional deviation of the gravitational redshift from the prediction by general relativity to be $\num{+0.19 \pm 2.48 e-5}$ at 1 sigma, improving the best previous test by Gravity Probe A by a factor~5.6. Moreover, we apply the exact same analysis to two almost circular Galileo satellites, in order to show the robustness of the method. By analyzing 899 days of circular Galileo satellites data we measure the fractional deviation of the gravitational redshift from the prediction by general relativity to be $\num{+0.29 \pm 2.00 e-2}$ at 1 sigma.
}

\section{Testing the Einstein equivalence principle with atomic clocks}
General Relativity is based on two fundamental hypothesis: the Einstein Equivalence Principle (EEP) and the Einstein field equations. Following Will \cite{Will1971a}, EEP can be divided into three \emph{sub-principles}: (i) the Weak Equivalence Principle: if any uncharged test body is placed at an initial event in space-time and given an initial velocity there, then its subsequent trajectory will be independent of its internal structure and composition; (ii) the Local Position Invariance (LPI): the outcome of any local non-gravitational test experiment is independent of where and when in the universe it is performed; (iii) and the Local Lorentz Invariance: the outcome of any local non-gravitational test experiment is independent of the velocity of the (freely falling) apparatus.

Tests of Lorentz Invariance have been performed using comparisons of atomic clocks onboard GPS satellites w.r.t.~ground clocks \cite{Wolf1997}, and more recently using an international network of optical clocks compared with optical fibres \cite{Delva2017e}. Test of Lorentz Invariance can be performed also in the Matter Sector \cite{Wolf2006,Hohensee2011,Pihan-LeBars2017,Sanner2019}. Test of LPI can be performed by searching for variations in the constants of Nature \cite{Uzan2011}. Several types of variation can occur depending on the particular theory: linear temporal drift \cite{Guena2012,Rosenband2008,Leefer2013,Godun2014,Huntemann2014}; spatial variation w.r.t.~the Sun gravitational potential \cite{Guena2012,Peil2013,Leefer2013,Ashby2007}; harmonic temporal variation \cite{VanTilburg2015,Hees2016}; and transients \cite{Derevianko2014,Wcislo2016,Roberts2017a,Wcislo2018}. Harmonic temporal variations and transients usually occurs when standard matter couples anormally to a field which can be linked to light dark matter.

Finally, LPI can be tested with a clock redshift experiment, where the gravitational redshift can be measured. It was observed for the first time in the Pound-Rebka-Snider experiment~\cite{Pound1960,Pound1959,Pound1959a,Pound1965}. In a clock redshift experiment, the fractional frequency difference $z=\Delta\nu / \nu$ between two identical clocks placed at different locations in a static gravitational field is measured. The EEP predicts $z = \Delta U / c^2$ for stationary clocks, where $\Delta U$ is the gravitational potential difference between the locations of the receiver and the emitter, and $c$ is the velocity of light in vacuum. A simple and convenient formalism to test the gravitational redshift is to introduce a new parameter $\alpha$ defined through~\cite{Will2014}:
\begin{equation}
	z = \frac{\Delta\nu}{\nu} = (1+\alpha) \frac{\Delta U}{c^2}
	\label{eq:redshift}
\end{equation}
with $\alpha$ vanishing when the EEP is valid. The most precise test of the gravitational redshift until recently has been realized with the Vessot-Levine rocket experiment in 1976, also named the Gravity Probe A (GP-A) experiment~\cite{Vessot1980,Vessot1979,Vessot1989}. The frequency differences between a space borne hydrogen maser clock and ground hydrogen masers were measured thanks to a continuous two-way microwave link. The gravitational redshift prediction was verified with \num{1.4e-4} uncertainty. A recent test with the clocks of Galileo 5 and 6 satellites improved this limit to \num{2.5e-5} uncertainty \cite{Delva2018c,Delva2015q,Herrmann2018}. The ACES experiment will test the gravitational redshift to around 2--\num{3e-6} accuracy \cite{Meynadier2018}. Furthermore, other projects like STE-QUEST propose to test the gravitational redshift at the level of $10^{-7}$. Finally, observations with the RadioAstron telescope reach today an accuracy \cite{Nunes2019} of \num{3 e-2}, and may reach \cite{Litvinov2017} an accuracy of the order of $10^{-5}$ in the future. 

In this article we report on the recent test of gravitational redshift using eccentric Galileo satellites \cite{Delva2018c}, where the fractional deviation of the gravitational redshift from the prediction by general relativity was reported to be $\num{+0.19 \pm 2.48 e-5}$, therefore compatible with general relativity. Moreover, we apply the same analysis to two circular Galileo satellites. Even if the accuracy of the test with circular satellites is 3 orders of magnitude smaller than the one with eccentric satellites, it is worth to verify that the exact same analysis lead to a robust estimation of uncertainties, and especially the ones due to systematics. Such a test with circular satellites has been performed with GPS in the TOPEX/POSEIDON Relativity Experiment \cite{Ashby2003}, reaching an accuracy of \num{2.2 e-2}, which is almost the same as our result with circular Galileo satellites.

\section{Test of the Gravitational redshift with Galileo eccentric satellites}

Galileo is the European Global Navigation Satellite System (GNSS). The space segment consists in 24 operational satellites and 6 spares satellites in medium Earth orbit, placed on three different orbital planes. There are actually 26 launched satellites. When the constellation will be completed, ten satellites will be spread evenly around each orbital plane. A global network of ground sensor stations is receiving signals from the Galileo satellites, and sending that information to control centres located in Europe. The control centres are computing the signals in order to find orbit solutions and synchronise the time signal of the satellites. The control centres are sending this information to the satellites via a global network of 5 up-link stations. Then, the satellites can send the relevant timing and positioning information to the receivers.

\begin{figure}
	\begin{center}
		\includegraphics[width=0.6\linewidth]{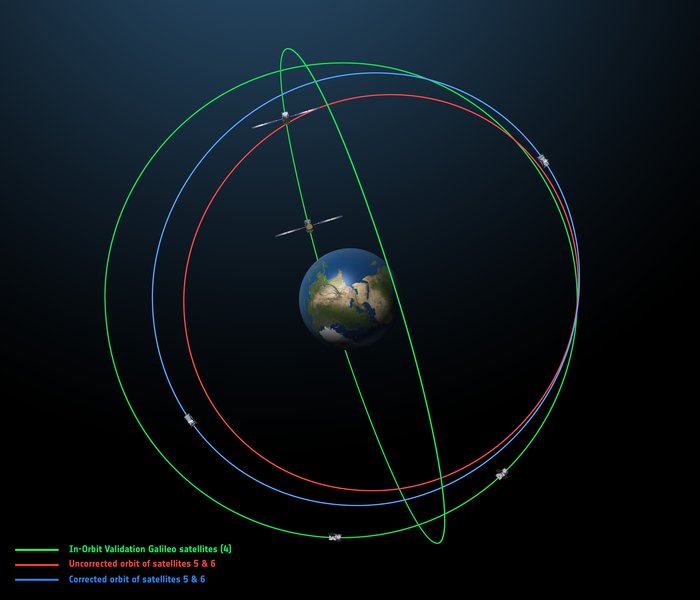}
        	\caption{\label{orbit} The original (in red) and corrected (in blue) orbits of the fifth and sixth Galileo satellites, along with that of the first four satellites (green) (Copyright ESA)}
	\end{center}
\end{figure}

ESA supports science done with Galileo through the GNSS Science Support Centre (GSSC: \url{gssc.esa.int}) and the GNSS Science Advisory Committee (GSAC). Today, more than 100 GNSS satellites have been launched, with a global coverage and continuous measurements. Their scientific use led to major contributions in Earth Science, Fundamental Physics, Metrology and many other fields. Before the second generation of Galileo satellites, an intermediate batch of 6 satellites will be launched between 2024 and 2026. These satellites will embark experimental instruments, which will be the occasion to perform new science with Galileo.

Galileo satellites GSAT0201 and GSAT0202 were launched on a Soyuz rocket on August, 2014 on the wrong orbit due to a technical problem. The anomaly occurred during the flight of Fregat, the launcher's fourth stage, 35 minutes after liftoff, leading to an anomaly in the orbital injection of the satellites \footnote{Arianespace Press Release: ``Soyuz Flight VS09: Independent Inquiry Board announces definitive conclusions concerning the Fregat upper stage anomaly'', October 08, 2014.}. Instead of a nominal circular orbit, the satellites were placed on a very eccentric orbit with 0.21 eccentricity (see Fig.\ref{orbit}, red orbit). The orbital error resulted from a temporary interruption of the joint hydrazine propellant supply to the thrusters, caused by freezing of the hydrazine, which resulted from the proximity of hydrazine and cold helium feed lines. This caused a loss of inertial reference and an error in the thrust orientation for the launcher's fourth stage. Soon after the launch, the European Space Agency searched for a suitable orbit for these satellites such that they could be operational \footnote{ESA Press Release: ``Galileo satellite recovered and transmitting navigation signals'', December 03, 2014.}. A total of 11 manoeuvres were performed to circularize the orbit, such that the satellite's Earth sensor can be used continuously, keeping its main antenna oriented towards Earth and allowing its navigation payload to be switched on. The final orbit of the two salvaged satellites is still very eccentric, with an eccentricity of 0.16 (see Fig.\ref{orbit}, blue orbit).

The Galileo satellites GSAT0201 and GSAT0202, with their high eccentricity, are perfect candidates for a test of the gravitational redshift. Indeed, an elliptic orbit induces a periodic modulation of the clock proper time at orbital frequency. For a Keplerian orbit it can be shown that the relation between proper time $\tau$ of the satellite clock and coordinate time $t$ is:
\begin{equation}
	\tau(t)=\left(1-\frac{3Gm}{2ac^2}\right)t-\frac{2\sqrt{Gma}}{c^2}e\sin{E(t)}+\text{Cst} \label{kepler}
\end{equation}
where $Gm$ is the gravitational parameter of the Earth, $a$, $e$ and $E$ are the semi-major axis, the eccentricity and the eccentric anomaly of the satellite, respectively, and $c$ is the velocity of light in vacuum. 

The first term in Eq.(\ref{kepler}), linear in time, results in a constant frequency shift between the satellite clock and a ground clock, which is $\approx \SI{40}{\micro\second\per\day}$ assuming a nominal 10.23 MHz frequency for the Passive Hydrogen Maser (PHM) onboard the satellite. However, each PHM clock is also affected by an intentional frequency offset ($\approx \SI{-6}{\micro\second\per\day}$) to this nominal frequency which explains an observed drift of $\approx \SI{34}{\micro\second\per\day}$. Additionally, after each activation the PHM clock retraces to the nominal frequency with an accuracy not better than $\pm \SI{0.18}{\micro\second\per\day}$. We account for this unknown frequency offset (together with the known $\approx \SI{34}{\micro\second\per\day}$) by removing from the measured clock bias a daily linear fit (DLF).

\begin{figure}
	\begin{center}
		\includegraphics[width=0.5\linewidth]{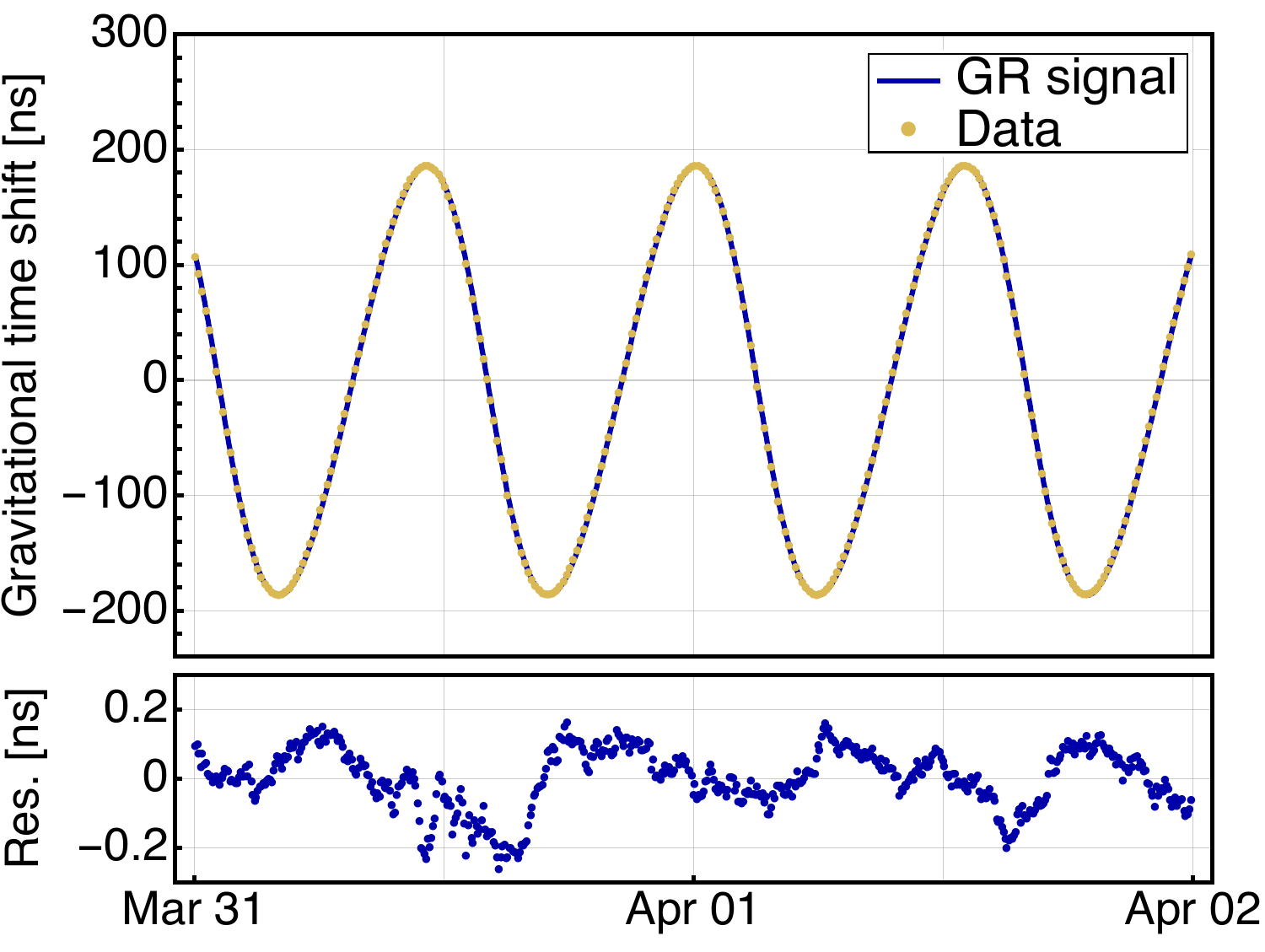}
        	\caption{\label{GR} GR prediction, clock data (after removal of a daily linear fit) and residuals are shown for 2 days from March 31st, 2016. The peak-to-peak effect is around \SI{0.4}{\micro\second}, therefore the model and systematic effects at orbital period should be controlled down to \SI{4}{\pico\second} in order to have a \num{1e-5} uncertainty on the LPI violation parameter~$\alpha$.}
	\end{center}
\end{figure}

The second term in Eq.(\ref{kepler}) is proportional to the eccentricity of the satellite and periodic with a period equal to the orbital period of the satellite. The amplitude of this term amounts to $\approx \SI{800}{\ns}$ peak-to-peak. To measure this periodic effect with a good accuracy, one needs a good stability of the clock over the orbital period, around \SI{13}{\hour}, which is the case of the Galileo clock, designed to be very stable over one day. Moreover, the nominal satellite lifetime is around 15~years, which allows to reduce the statistical uncertainty of the gravitational redshift test, and the satellites are permanently monitored by several ground receivers.

The orbit of the Galileo satellites is not Keplerian. Therefore we cannot use Eq.(\ref{kepler}) to calculate the total redshift. Instead, one has to use the full formula from GR, which can be found in \cite{Delva2018c}. As input to the full proper time to coordinate time transformation, we use an orbit and clock solution generated by ESA's Navigation Support Office, located at the European Space Operations Centre (ESOC). The proper time to coordinate time transformation equation can be split into different contributions: the periodic gravitational redshift amounts to \SI{400}{\nano\second} peak-to-peak (see Fig.\ref{GR}); the Earth flatness leads to a \SI{40}{\pico\second} peak-to-peak periodic effect at twice the orbital frequency; finally, tidal effects from the Moon and the Sun lead to a periodic signal of around \SI{12}{\pico\second} peak-to-peak, higher than the uncertainty goal of the experiment which is \SI{4}{\ps}.

The data analysis is done in three steps. First, we fit a model for the stochastic noise to the corrected clock bias residuals. In a second step, we fit our theoretical model to the corrected clock bias by using a Monte Carlo approach, using the stochastic noise model estimated in the first step. This gives us the fitted value for the LPI violation $\alpha$ as well as an estimation of its statistical uncertainty. In a third step, we estimate the systematic uncertainty by considering the main sources of systematics: effects of magnetic field, of temperature and mismodelling of the orbital motion of the satellites. These three steps are described in details in \cite{Delva2018c}.

Finally, by analysing 1008~days of data from the two eccentric Galileo satellites, GSAT0201 and GSAT0202, and through a careful analysis of systematic effects, we were able to improve the gravitational redshift test done by GP-A in 1976 by a factor 5.6, down to $\alpha = \num{+0.19 \pm 2.48 e-5}$. The different sources of uncertainty are given in Table~\ref{tab:result}.

\section{Test of the Gravitational redshift with Galileo circular satellites}

\begin{figure}
	\begin{center}
		\includegraphics[width=0.7\linewidth]{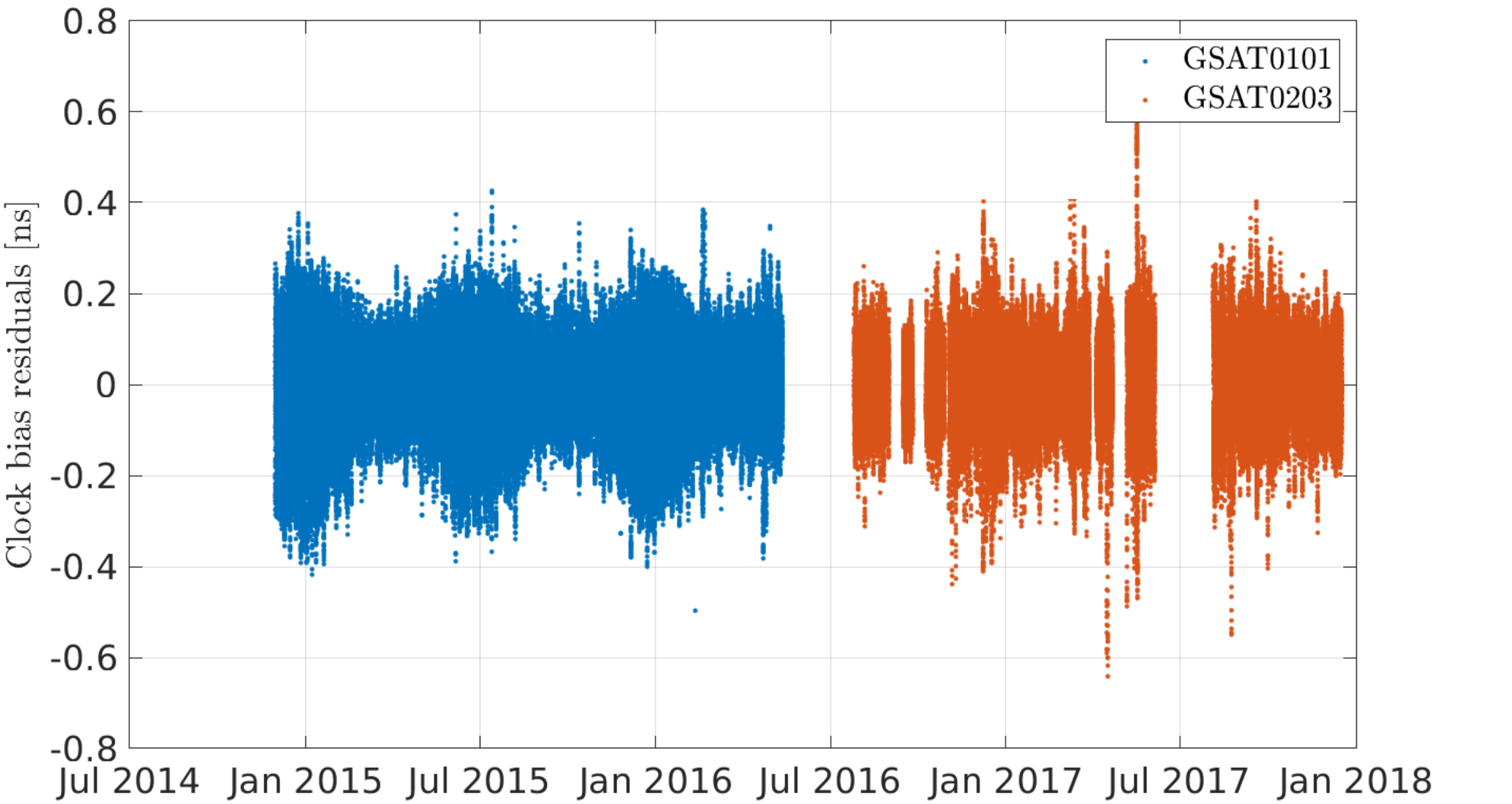}
        	\caption{\label{clk} Clock bias pre-fit residuals are obtained by removing from the raw clock bias a daily linear fit. Here only the times taken into account in the analysis are shown.}
	\end{center}
\end{figure}

Now we apply exactly the same data analysis to two chosen circular satellites. The master clock on board the Galileo satellites may change over time due to maintenance routine. There are two PHM clocks as well as two rubidium clocks (RAFS) on board each of the satellites. The RAFS is usually ten times noisier than than the PHM. The circular satellites have been chosen such that one of the PHM clock was operational for more than one year. We did not use the RAFS clock. We choose one satellite from the IOV (In-Orbit Validation) generation, GSAT0101, which was launched in October 21, 2011. Moreover, we choose another satellite from the FOC (Full Operational Capability) generation, GSAT0203, which was launched in March 27, 2015. The pre-fit residuals of the clock bias from these two satellites are shown in Fig.\ref{clk}. The data analysis contains 518 days of data from GSAT0101 and 381 days of data from GSAT0203, spanning from November 2014 to December 2017.

The different sources of uncertainty are given in Table~\ref{tab:result}. When we quadratically add the statistical and systematic uncertainties due to each considered error source, we obtain for the LPI violation parameter $\alpha = \num{0.71 \pm 1.89 e-2}$ for GSAT0101 and $\alpha = \num{-5.31 \pm 4.86 e-2}$ for GSAT0203. Finally, we combine the data from both circular satellites using a global Monte-Carlo least square analysis, by applying a weight inversely proportional to their orbit uncertainty given in Table~\ref{tab:result}, as we did for the eccentric satellites. We obtain for the LPI violation parameter $\alpha = \num{0.29 \pm 2.00 e-2}$. The aacuracy of the test is slightly degraded in the combination compared to GSAT0101 alone, because of the high orbit uncertainty of GSAT0203, however the bias is closer to zero in the combined solution.

\section{conclusion}

To conclude, we reported the use of eccentric Galileo satellites to perform the best test of the gravitational redshift to date \cite{Delva2018c}. By analysing 1008~days of data from the two eccentric Galileo satellites, GSAT0201 and GSAT0202, and through a careful analysis of systematic effects, we were able to improve the gravitational redshift test done by GP-A in 1976 by a factor 5.6, down to $\alpha = \num{+0.19 \pm 2.48 e-5}$. Our result is at the lower edge of the predicted sensitivity in~\cite{Delva2015q}. Moreover, we applied the same analysis to two circular Galileo satellites, GSAT0101 and GSAT0203, obtaining $\alpha = \num{0.29 \pm 2.00 e-2}$. This new result shows the robustness of the uncertainty analysis we developed for the eccentric satellites. Finally, the ACES experiment will improve this result by one order of magnitude in 2020. Ground optical clock comparisons might also improve our result in the coming years.

%{\renewcommand{\arraystretch}{1}
%\setlength{\tabcolsep}{1pt}
\begin{table}[h]
	\centering
	{\small
	\begin{tabular}[t]{lcccccc}
		\hline\hline
		& LPI  & Total  & Stat. & Orbit & Temperature & Magnetic field \\
		& violation & uncertainty &  unc. &  unc. &  unc. & tot. unc. \\\hline
%		& & & (CLK fit) & unc. & (SLR fit) & (nominal) & (nominal) \\
		\rule[-0.2cm]{0cm}{0.6cm}Eccentric & [$\times10^{-5}$] & [$\times10^{-5}$] & [$\times10^{-5}$]  & [$\times10^{-5}$] & [$\times10^{-5}$] & (X/Y/Z) [$\times10^{-5}$] \\\hline
		GSAT0201 & $-0.77$ & $2.73$ & $1.48$ & $1.09$ & $0.59$ & $1.93$ \ $(0.52,-0.36,1.82)$  \\
		GSAT0202 & $6.75$ & $5.62$ & $1.41$ & $5.09$ & $0.13$ & $1.92$ \ $(-0.07,0.58,1.83)$ \\
%		\rule[-0.2cm]{0cm}{0.6cm} \bf{Combined} & $\bm{0.24}$ & $\bm{2.50}$ & $1.29$ & $1.01$ & $0.10$ & $1.88$ \ $(0.44 / -0.22 / 1.82)$ \\\hline
%		\rule[-0.2cm]{0cm}{0.6cm} \bf{Combined} & $\bm{-0.02}$ & $\bm{2.48}$ & $1.30$ & $0.75$ & $0.54$ & $1.90$ \ $(0.47 / -0.28 / 1.82)$ \\\hline
		Combined & $0.19$ & $2.48$ & $1.32$ & $0.70$ & $0.55$ & $1.91$ \ $(0.48,-0.29,1.82)$ \\\hline
		\rule[-0.2cm]{0cm}{0.6cm}Circular & [$\times10^{-2}$] & [$\times10^{-2}$] & [$\times10^{-2}$]  & [$\times10^{-2}$] & [$\times10^{-2}$] & (X/Y/Z) [$\times10^{-2}$] \\\hline
		GSAT0101 & $0.71$ & $1.89$ & $0.57$ & $1.39$ & $0.07$ & $1.15$ \ $(-0.07,1.15,-0.01)$\\
		GSAT0203 & $-5.31$ & $4.86$ & $0.90$ & $4.58$ & $0.07$ & $1.35$ \ $(0.05,1.35,0.00)$\\
		Combined & $0.29$ & $2.00$ & $0.55$ & $1.53$ & $0.07$ & $1.16$ \ $(-0.06,1.16,-0.01)$\\
		\hline\hline
	\end{tabular}
	}
	\caption{Final result of the EEP/LPI test. Each row is a separate output from fits of each satellite data and models, and a combined fit of 2 satellites data and models (either eccentric or circular). The uncertainties due to systematic effects are evaluated independently from the clock data, and are the result of individual fits of models of systematics (for temperature and magnetic field), and a fit of SLR residuals for the orbit uncertainty, thus giving an upper limit of each effect. A single value is computed for the magnetic field uncertainty by summing in quadrature the X/Y/Z values. The total uncertainty column, for each row, is derived from the quadratic sum of the individual uncertainties to the right.}
	\label{tab:result}
\end{table}

\section*{Acknowledgments}
We acknowledge financial support from ESA within the GREAT (Galileo gravitational Redshift Experiment with eccentric sATellites) project, from Paris Observatory/GPhys specific action.
\section*{References}

%\bibliography{moriond}

\end{document}